# Design and development of the PRIMAger instrument for PRIMA

Anne Costille *[a], Fabrice Madec[a], Johan Floriot[a], Tony Pamplona[a], Thomas Prouvé[b], Laure Ciesla[a], Marc Sauvage[c], Willem Jellema[d], Lorenza Ferrari[d], Jochem Baselmans[d], Florent Canourgues[f], Renaud Goullioud[e], Carole Tucker[g], C. Darren Dowell[e]
[a]Aix Marseille Univ, CNRS, CNES, LAM, Marseille, France; [b]Univ. Grenoble Alpes, CEA/IRIG/DSBT, Grenoble 38000, France, [c]Université Paris-Saclay, Université Paris Cité, CEA, CNRS, AIM, 91191, Gif-sur-Yvette, France,[d]SRON Space Research Organization Netherlands (The Netherlands), [e]Jet Propulsion Laboratory, California Institute of Technology, **[f]**Ctr. National d'Études Spatiales (France), [g]Celtic Terahertz Technology, [f]Delft University of Technology, Delft (The Netherlands)

**ABSTRACT**

PRIMA (PRobe far-Infrared Mission for Astrophysics) is an infrared space observatory designed to operate over the next decade. Equipped with a 1.8 m telescope actively cooled to 4.5 K, the project is currently in the concept study phase as part of the NASA Probe Explorers program. PRIMA will carry two instruments: FIRESS and PRIMAger. PRIMAger will offer two observation modes: hyperspectral imaging and polarimetric imaging. Both channels will use ultra-sensitive kinetic inductance detector (KID) arrays, covering wavelengths from mid-infrared to far-infrared. A European consortium will develop the cold unit of the PRIMAger instrument while the Readout Electronics and Continuous Adiabatic Demagnetization Refrigerator (CADR) will be provided by Goddard Space Flight Center (GSFC) and shared between the 2 instruments. The hyperspectral mode will cover the 24–84 µm range with a spectral resolution of R ≥ 8, ensured by a linear variable filter placed in front of the detector. The polarimetric mode will provide polarimetric imaging in four broad bands from 92 to 235 µm, using KIDs with three polarization orientations. The main structure of the instrument is the opto-mechanical assembly, cooled at 1K. It supports at first the two focal plane array assemblies incorporating 0.125K detector stages. In addition, the main structure is providing the interfaces for the 0.125K thermal link, the readout cables and the filters. This paper presents the technical requirements and expected performance of the instrument, as well as the associated design of the sub-systems ensuring compliance. A major challenge will be in-flight performance verification. We will also detail the development strategy, including the planned verification and test program.

**Keywords:** Far-Infrared, Kids Detector, Polarimetry, Hyperspectral, PRIMA

## 1. INTRODUCTION

The PRIMAger Cold Unit instrument is one of the two science instruments foreseen within the PRobe far-Infrared Mission for Astrophysics (PRIMA) [1], developed by NASA for the next decade. The project is currently under a selection phase to offer the instrument to the community by 2032 pending a selection decision from NASA. PRIMA benefits from a 1.8m telescope actively cooled to 4.5K. PRIMAger will be the infrared camera of the observatory and will provide coverage from mid-infrared wavelengths to far-infrared wavelengths from 24 to 264µm thanks to the use of ultra-sensible kinetic inductance detector (KID) [4;5]. PRIMAger will provide two imaging modes: the hyperspectral mode will cover the 24-84 µm wavelength range with a spectral resolution R≥8, while the Polarimetric mode will provide polarimetric imaging in 4 bands with R~4.5, from 92 to 235µm. The observational capabilities will be used to answer fundamental astrophysical questions as described in [2].
PRIMAger, the cold unit, is being developed by an international collaboration held by CNES with the contribution of French institutes, Laboratoire d'Astrophysique de Marseille and CEA, the Netherlands institute for Space Research (SRON), Delft University of Technlogy (TU Delft), and the Cardiff University (UK) in Europe. The Jet Propulsion Laboratory (JPL) and Goddard Space Flight Center (GSFC) in the USA are managing the overall PRIMA project and provide to PRIMAger the needed interfaces to make the instrument work. In this article, we will describe first the main requirements of the instrument, and the architecture and design that will allow to provide the expected performance. Then we will present the development strategy, including the planned verification and test program to verify the performance of the instrument prior delivery to the payload.

*anne.costille@lam.fr; phone +33 (0)4 12 23 18 63; www.lam.fr

## 2. MAIN TECHNICAL REQUIREMENTS

PRIMAger will be part of the the PRIMA system that features a diffraction-limited, afocal three-mirror telescope with a 1.8 m primary mirror (TM1), which acts as the aperture stop for the combined telescope-instrument system (FIRESS or PRIMAger) (Figure 1-Left). The telescope targets two off-axis sky fields—one for each instrument—routed via collimating tertiary mirrors (TM3) and folding mirrors. Its exit pupil, located on the scanning-capable Beam Steering Mirror (BSM), delivers 50 mm collimated beams to PRIMAger with a stabilized line of sight (LoS). Both the telescope and PRIMAger operate in the far-infrared range (24–264 µm).

PRIMAger's scanned field of view (FoV) spans 36.6′ × 27′ (Figure 1-Right), positioned at an off-axis angle of (-0.425°; 0.565°) relative to the telescope's central reference frame (XTFov, YTFov, ZTFov). As a dual-channel FIR imager, PRIMAger simultaneously covers the 24–264 µm range using kinetic inductance detectors (KIDs). Its hyperspectral (PHI) and polarimetric (PPI) focal planes share the same FoV but are spatially offset by 4.7′ on the sky (Figure 1-Right). The BSM enables rapid 2D scanning within the telescope's 37′×27′ FoV, and larger sky areas can be mapped by combining this with the observatory's motion. While each detector images a unique portion of the FoV, the scanning strategy ensures overlapping sky coverage by multiple detectors.

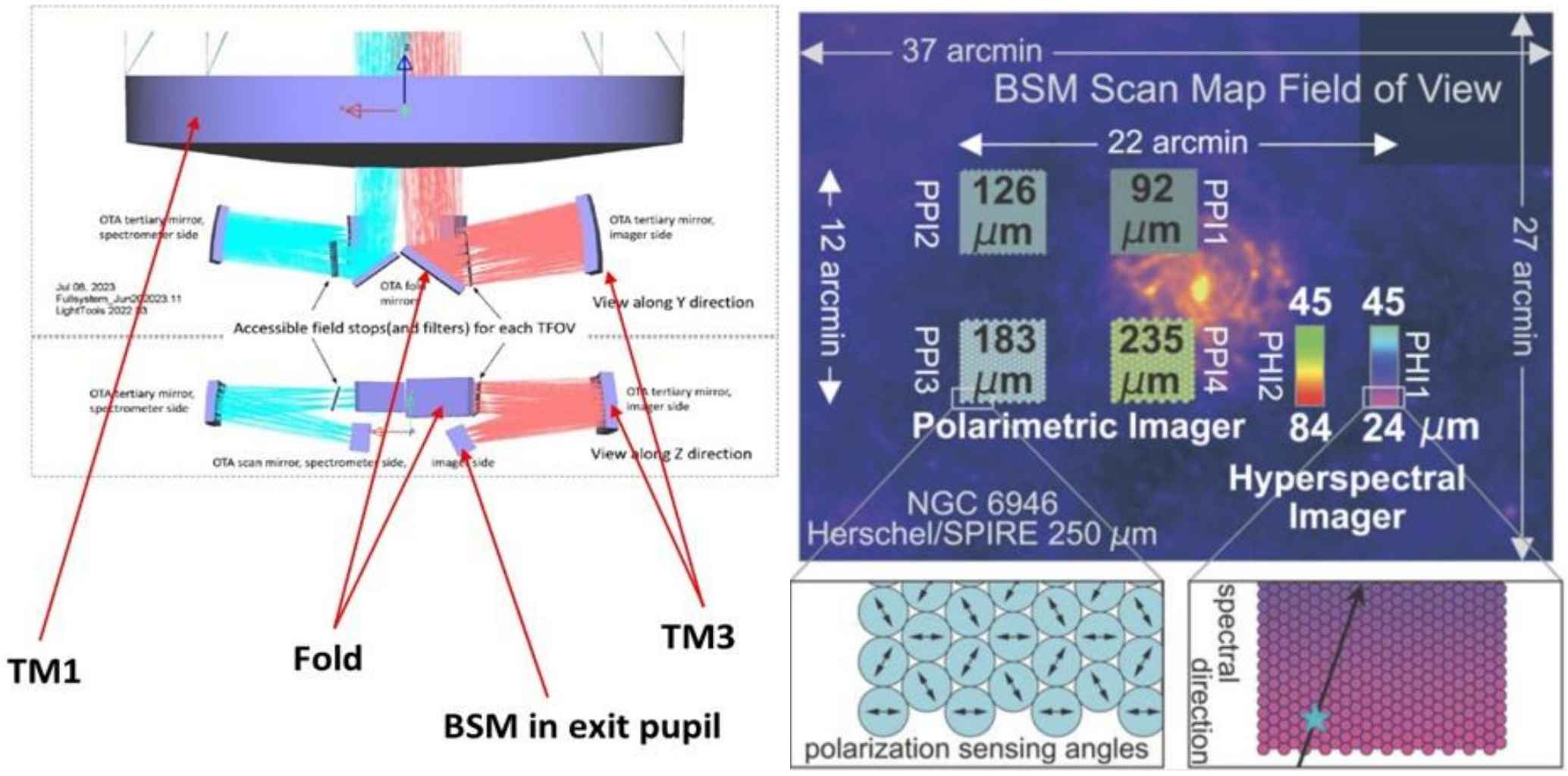

*Figure 1. Left: optical path of the PRIMA telescope. Right: FoV defined for PRIMAger instrument.*

### 2.1 PRIMAger requirements

The requirements of PRIMAger cold unit are provided and agreed by JPL. The following table summarizes the main performance requirements of the instrument from the flow down of the science requirements. The instrument will provide 2 observation modes:

- An hyperspectral imaging mode over two wavelength ranges:
- A polarimetric imaging mode for 3 linear polarization angles.

| **Parameter** | **Values (requirement & design value)** | |
|---|---|---|
| | Hyperspectral PHI | Polarimetric PPI |
| Modes | 2 bands | 4 bands with 3 linear polarization |
| Resolution R = lambda/dLambda | 8 | 4 |
| Spectral range | 24 to 45 µm, this band will be called hereafter PHI1<br>45 to 84 µm this band will be called hereafter PHI2 | 81 to 103 µm, this band will be called PPI1<br>110 to 142 µm, this band will be called PPI2 |
| | | 160 to 206 µm, this band will be called PPI3<br>206 to 264 µm, this band will be called PPI4 |
| Transmission | ≥ 0.14 | ≥ 0.21 |
| Pixel sampling | ≤ Lambda/d | ≤ Lambda/d |
| F/# | 21 | 12 |
| Field of view (arcmin) | 1.4 x 3.9 for each PHI | 4.2 x 4.2 for each PPI |
| Sensitivity | $\nu * \sigma(F_\nu) \leq$ 5.5E-17 $(W/m^2)*sec^{1/2}$ (1σ, statistical) for a point source located randomly within the field of view, for assumed background intensity of 10.5E-20, 5.2E-20 $W/m^2/Hz/sr$ at 35, 63 µm | 2.9E-22 $(W/m^2/Hz/sr)*sec^{1/2}$ (1σ, statistical) for total intensity and 4.2E-22 $(W/m^2/Hz/sr)*sec^{1/2}$ (1σ, statistical) for polarized intensity, for assumed background total intensity of 12.1E-20, 9.1E-20 $W/m^2/Hz/sr$ at $\lambda$ = 92, 235 µm |
| Design Limit Load | 20g at system level<br>100g at detector level | |
| Mass | < 51.8kg | |
| Cooling power with T2 = 4.5K | ≤ 104µW at T1 = 1K<br>≤ 2.2µW at T0 = 0.125K | |

## 2.2 Instrument architecture

The instrument architecture has been defined so that we can define clear interfaces between the different parts and also clear responsibility in the delivery of a self-contained parts of the instrument. The PRIMAger product tree has then been broken into 5 sub-systems of the instrument distributed as followed:

- Focal Plane Module Assembly (FPMA) which contains both PHI and PPI detectors.
- Mechanical-Optical-Thermal Assembly (MOTA) which contents the structure and the optical mirrors
- Thermal Link Assembly (TLA) which allows the cool down of the T0 stage instrument
- Electrical Link Assembly (ELA) which includes both coax harness and temperatures sensors
- MOTA Filter Assembly (FA) for all filters within MOTA. The other filters are directly part of FPMA.

In addition we have defined Ground Support Equipment (GSE):

- Mechanical and electrical GSE.
- Thermal Vaccuum Optical Ground Support Equipment (TVOGSE) that includes all the elements developed for the performance verification of the instrument.

Each sub-assembly is under the responsibility of an institute and are assembled to form the PRIMAger Cold Unit that is delivered to JPL.

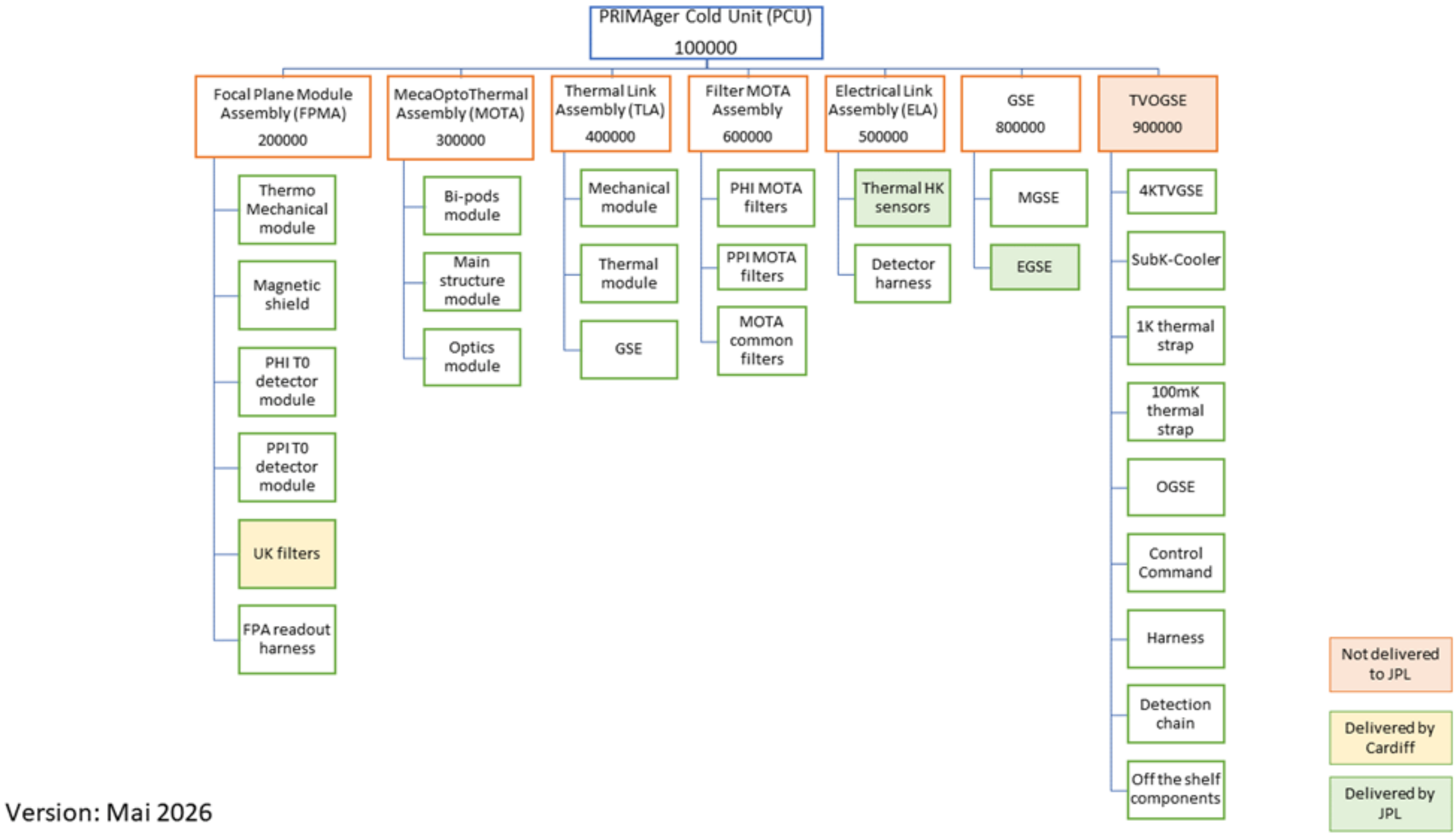

*Figure 2. PRIMAger product breakdown structure*

## 2.3 Instrument functional description

Figure 3 shows the functional diagram for PRIMAger. The instrument receives light from PRIMA payload through the optical system and distributes it into the two detector focal planes: one for PPI and one for PHI. A set of filters distributed among the optical path through the detectors allows rejection of the out-of-band wavelengths.

The instrument is directly in interface with the PRIMA payload metering structure at 4.5K thanks to the bipods that supports the entire 1K structure, in which the optics and the detector focal plane are located. The structure provides the mechanical interface required by the instrument while the bipods ensure low thermal conductivity to the PRIMAger 1K structure. The structure is at 1K to limit thermal background for the detectors, while the detectors are heat sinked at 0.125K thanks to an internal thermal link at 0.125K. . The focal planes are surrounded by magnetic shielding to allow rejection of the magnetic field. Two harnesses are provided: one for the readout electronics, and one for the thermal sensors. Both are part of the Electrical Link Assembly.

The PRIMAger Cold Unit is a passive instrument as there is no mechanism and no electronics in the instrument. It interfaces with two systems developed by GSFC, shared with FIRESS instrument, to ensure operation and performance:

- A readout electronics box, under Goddard Space Flight Center, GSFC
- A cryogenic CADR system (Continuous Adiabatic Demagnetization Refrigerator), under the responsibility of Goddard Space Flight Center, GSFC

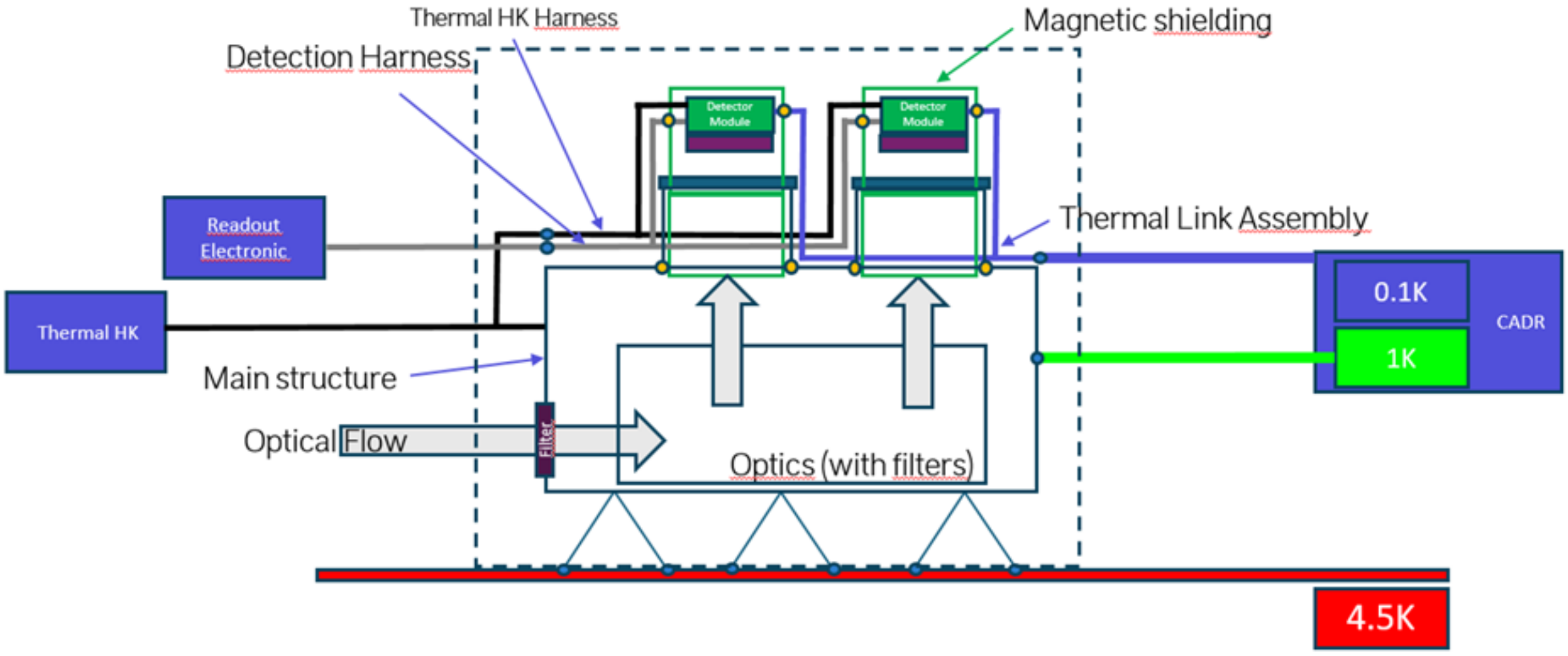

*Figure 3. Functional diagram of PRIMAger.*

## 3. PRIMAGER DESCRIPTION

PRIMAger is made of several sub-systems that are described in the next sections. PRIMAger has two optical paths to answer to the scientific objectives of the instrument:

- PRIMAger Hyperspectral Imager (PHI)
- PRIMAger Polarimetric Imager (PPI)

PRIMAger consists of several parts: the 1K Mechanical-Optical-Thermal Assembly (MOTA) interfaces to the payload 4.5K Aft Mechanical Structure, the two focal planes module assemblies (FPMA), that includes the 0.125K focal planes modules, and the suspended Thermal Link Assembly (TLA) that supports the straps providing cooling power to the focal plane modules. The MOTA aluminum box provides a 1K environment for the PRIMAger optics and the long and short pass filters. The FPMAs provide optimal environmental conditions for the detectors: each FPMA consists of a housing at 1K that interfaces to MOTA, two layers of magnetic shielding, a Kevlar cord suspension system to thermally isolate the 0.125K module from the 1K housing and a focal plane module at 0.125K containing the KID detector arrays in a light tight box.

The total mass of the instrument is equal to 51.8kg, including design contingency, for a volume of 700x670x700mm. Figure 4 shows the current design of the instrument.

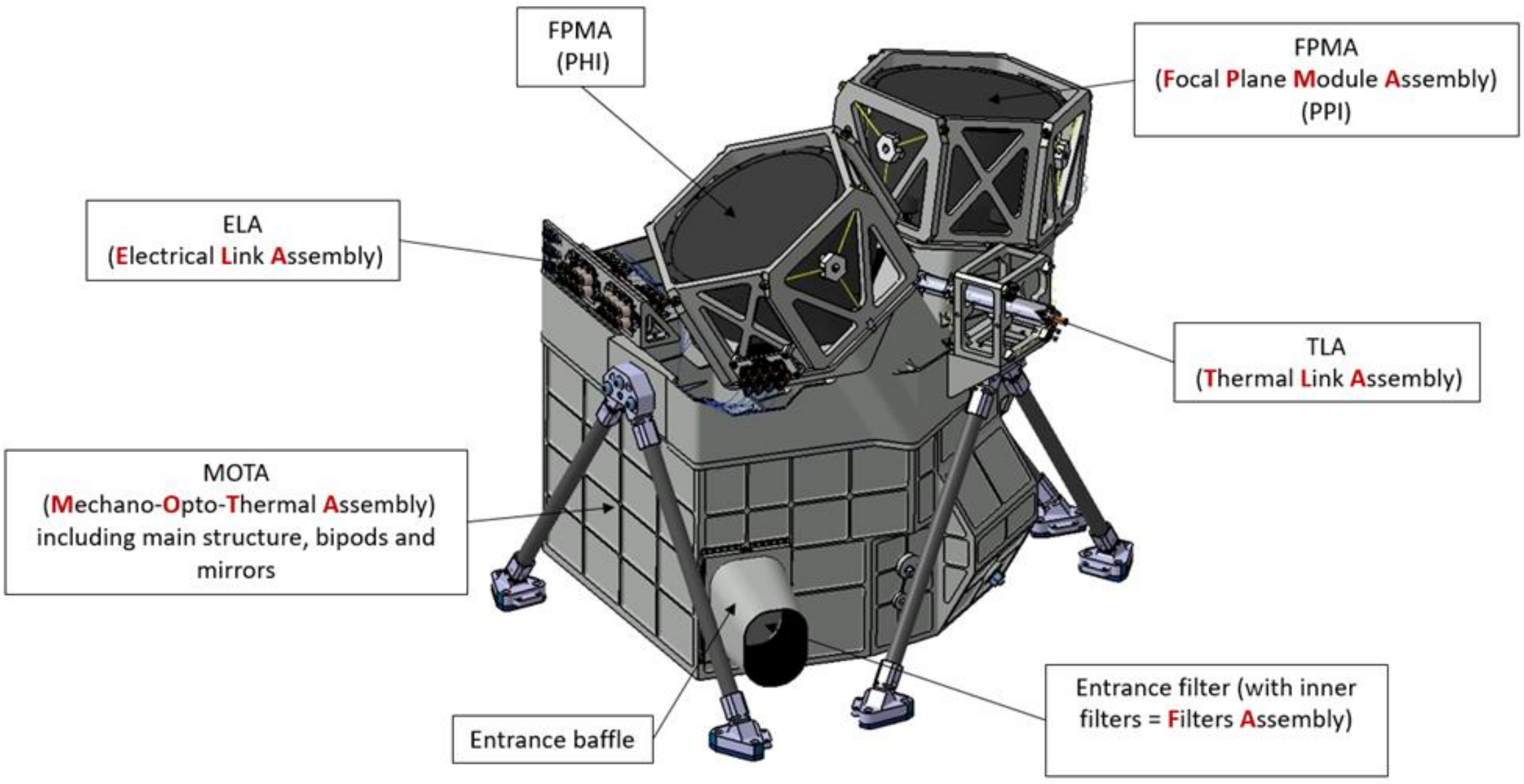

*Figure 4. PRIMAger instrument and its sub-systems.*

The optical layout of PRIMAger is shown in Figure 5-Left. The instrument interfaces with the 1.8m telescope thanks to an exit pupil placed on the beam steering mirror, which acts as an aperture stop (50mm diameter) of PRIMAger. At the entrance of PRIMAger box, an entrance filter is set to reject out-of-band-light. Then the light hits a common mirror (PM1) that allows the spatial separation of the hyperspectral and polarimetric fields of view. From this mirror, the beams of the PHI and PPI imagers follow different paths with similar optical architectures. Each optical path is composed of fields mirrors (PM2) on the focus position of PM1. An intermediate pupil plane is then available for a cold stop and quasioptical filters. One additional spherical mirror (PM3) is present and then a nearly flat mirror (PM4) is set just in front of the detector planes to define the beam at the entrance of the detector modules. The PHI imager is F/21 and the PPI imager is F/12. The PHI image quality is quasi-diffraction-limited, while the PPI image quality is diffraction-limited. All the parameters of the optical design have been defined to obtain a spatial resolution of λ/D on the pixel. The full description and analysis of the optical design is provided in [3].

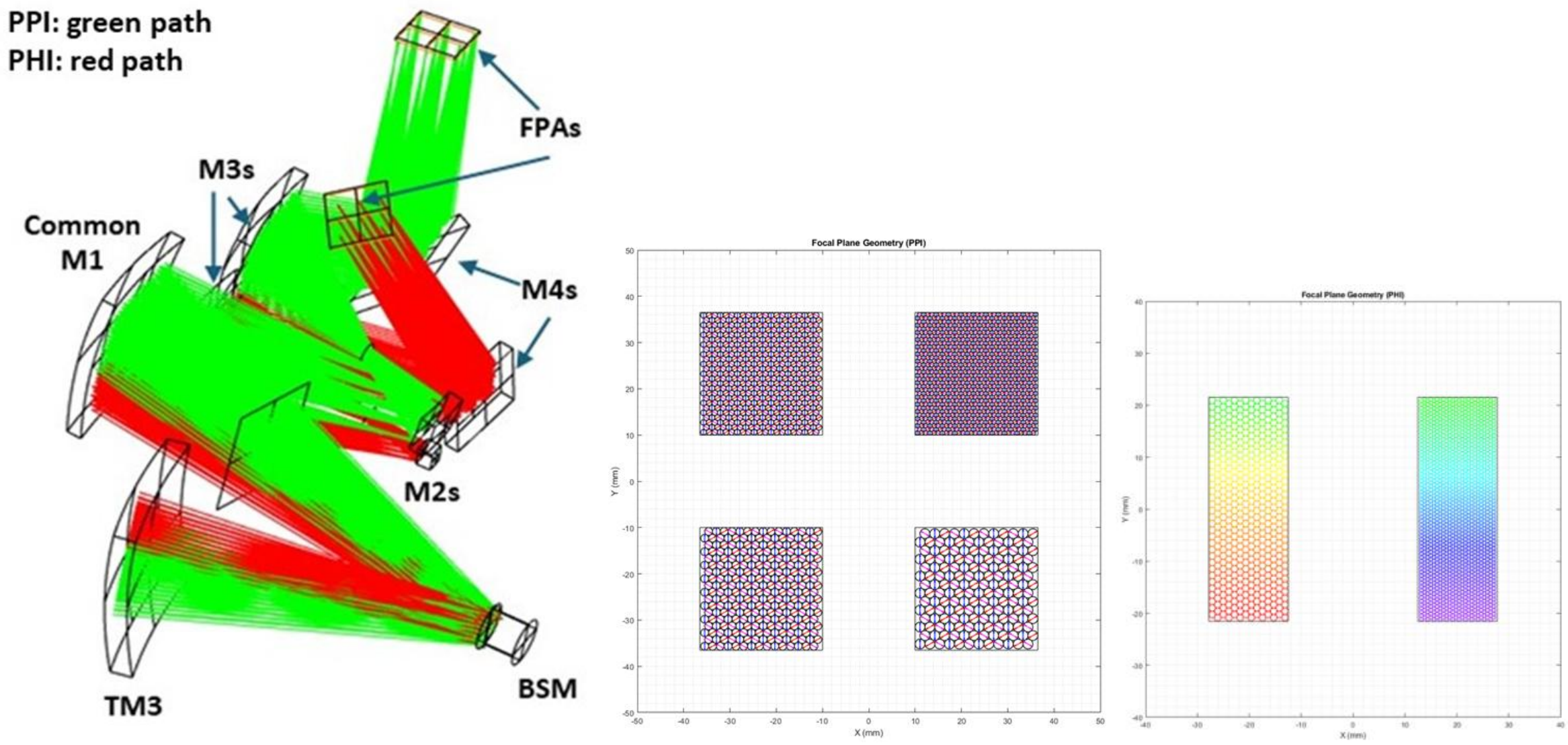

*Figure 5. Left: Layouts of PPI and PHI inside PRIMAger. Green path: PPI, Red path: PHI. Middle: PPI Detector arrangement. Right: PHI detector arrangement.*

The optical design allows to provide light to the two focal plane assemblies and their associated detectors. Figure 5 middle & right shows the arrangement chosen for the detector array in the FoV. PRIMAger will provide:

- 4 PPI detectors of 4.2'x4.2', separated with 3.2' clearance (20mm mechanical clearance). This will lead to a total FoV of PPI of 11.6'x11.6;
- 2 PHI detectors of 1.4'x3.9', separated by 2.3' clearance (25mm mechanical clearance). This will lead to a total FoV of PHI of 5'x3.9'.
- Both FoV are separated with 0.08°, edge-to-edge FoV separation.

## 3.1 MOTA description

MOTA (Mechanical-Optical-Thermal assembly) is the main structure of PRIMAger that holds the optics and the detector focal planes. MOTA is an aluminum structure. It is developed by the Laboratoire d'Astrophysique de Marseille (LAM), responsible for the design and fabrication of MOTA, including the PRIMAger optics and their mounting, integration, alignment and verification. A detailed description of MOTA is provided in [3].
MOTA features an aluminum box cooled to 1K. The structure is both light-weighted and reinforced to optimize mass to stiffness ratio and respect the related requirements. The chosen material is aluminum 6061 T6, typically used for cryogenic applications. The design of MOTA is trade-off to find the best solution. For phase A, the baseline design consists of a box to which the aluminum mirrors are attached to. Each mirror is isostatically mounted on the box via three fixed points allowing accurate positioning, alignment and stability. The material used to manufacture these mirrors is the same as that of the main box ensuring a homothety during cool-down maintaining optical alignment and performance. An alternative design is also being study to perform a trade-off analysis for the start of the phase B. It consists of a main bench to which all mirrors are attached thanks to a mechanical support that allows it alignment. To maintain the instrument on the payload, the design is based on a kinematic system of three bipods allowing its thermal decoupling from the telescope, particularly the conductive thermal insulation between the 4.5-K telescope platform and the 1-K PRIMAger main box. This optomechanical design is compatible with typical space environment conditions: cryogenic operating temperatures leading to cooling down from 293K to 0.1K and launch vibration levels of 20g.

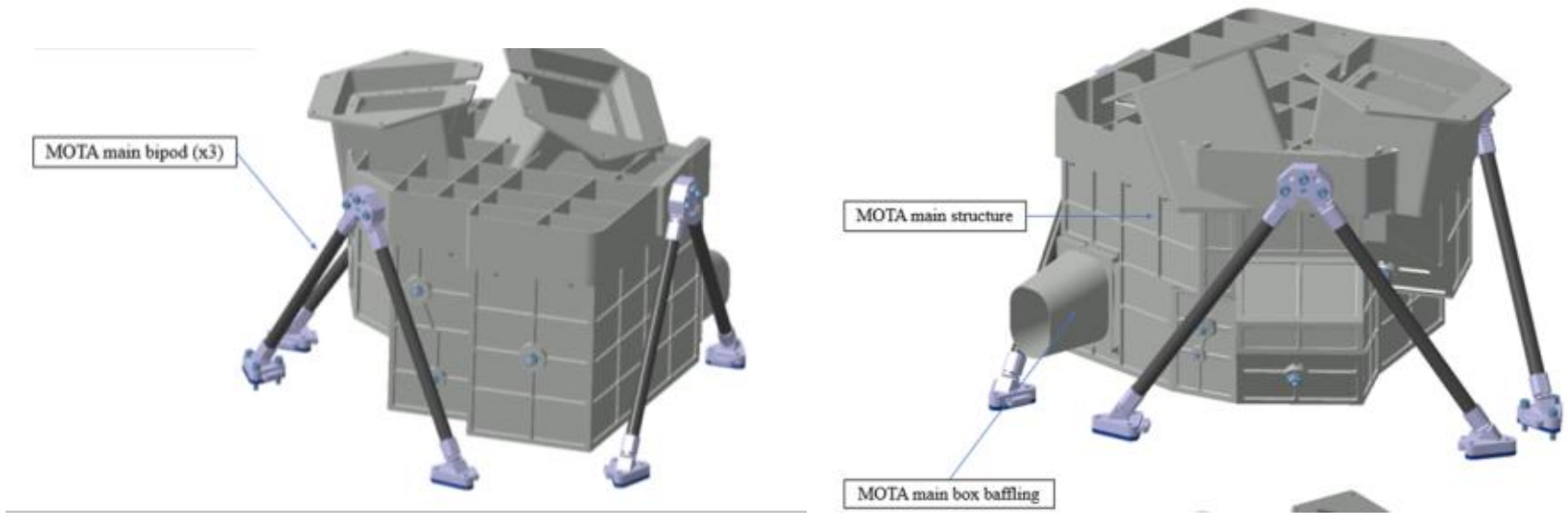

*Figure 6. MOTA 3D overview*

MOTA provides mechanical interfaces to all PCU sub-systems and allows for the routing of the necessary harnesses. It also provides the optical interfaces to the FPMA. MOTA is also the primary interface with PRIMA payload for optical and mechanical interfaces.

## 3.2 FPMA description

SRON is responsible for the design, fabrication, integration and test of the Focal Plane Module Assembly (FPMA), including the KID arrays with lenslet arrays, the focal plane module including the detectors, the suspension system and the magnetic shields.
The Phase A design of the PRIMAger PPI and PHI FPMAs employs KID detector arrays operating at T0 ~ 125 mK to realize background-limited detection between 24-264 µm.

- Polarimetric imaging in PPI's four R=4 bands between 81-264 µm is realized by the use of linearly polarized antenna-coupled KID detectors, with 3 pixel types sensitive to polarization orientations at 0° and +/-60°. This allows Stokes parameter (I, Q, U) maps to be derived from intensity maps in 3 polarizations, combining telescope slews and beam-scanning mirror scans to map the targeted region on the sky with all 4 PPI detectors[4,6,7].

- Hyperspectral imaging in PHI's 2 bands between 24-85 µm is realized by placing Linear Variable Filters (LVFs) in front of 2-D arrays of absorber-coupled KID detectors [5], producing R≥7 spectra along the length direction of each detector array (current design aims for R=8), with the perpendicular direction of the array providing additional "spatial" pixels to increase mapping speed. Because each instantaneous sky position is measured in one sub-band of the PHI wavelength range, spectral mapping requires a combination of telescope slews and/or beam-scanning mirror scans to observe the targeted region on the sky over the full spectral range of both PHI detectors.

The following table summarizes the main characteristics of the FPMA.

| **Requirement** | **PPI** | | | **PHI** | | |
|---|---|---|---|---|---|---|
| | PHI1 | PHI2 | PPI1 | PPI2 | PPI3 | PPI4 |
| Array size (nb of pixels H x V) | 24x61 | 14x34 | 29x34 | 21x25 | 15x18 | 12x14 |
| Number of pixels | 1464 | 476 | 986 | 525 | 270 | 168 |
| Pixel pitch (mm) | 0.7 | 1.258 | 0.896 | 1.216 | 1.685 | 2.12 |
| Instantaneous FoV(arcmin) | 1.3x3.9 | 1.4x3.9 | 4.2x4.2 | 4.2x4.2 | 4.2x4.2 | 4.2x4.2 |

The tiling of the PPI and PHI detectors in the focal plane is illustrated in Figure 7.

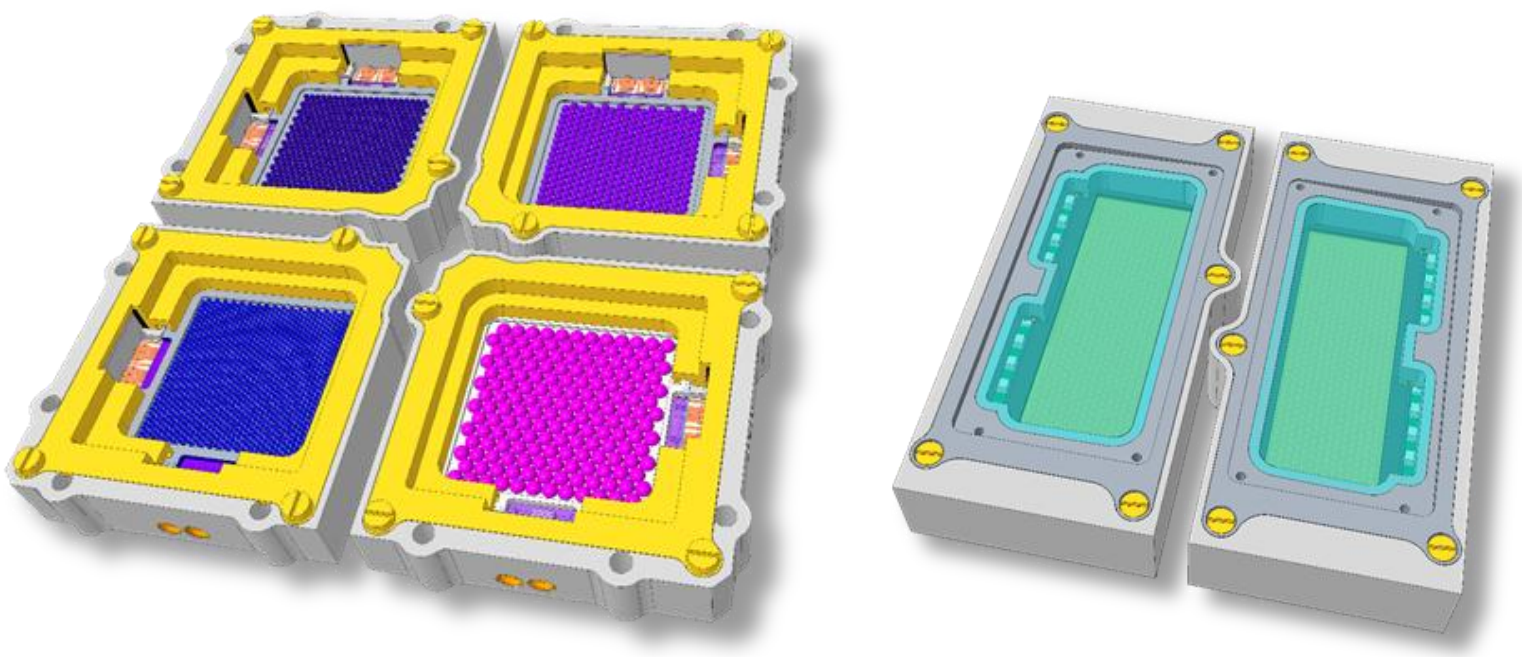

*Figure 7. Preliminary concept for the tiling of (left) 4 PPI detectors and (right) 2 PHI detectors.*

With the detectors operating at T0 ~ 125 mK and the FPMA's mechanical interface to the PRIMAger opto-mechanical assembly (MOTA) at T1 ~ 1 K, a thermal suspension is needed that combines high strength and stiffness (to survive launch vibration and decouple the first eigenfrequency of the FPMA from that of the MOTA) and high thermal isolation (low thermal conduction) between the T1 and T0 stages. The baseline FPMA concept applies a suspension design that is scaled from the focal plane assembly in Athena-XIFU. A quasi-isostatic construction using pre-tensioned Kevlar cords provides the required strength, stiffness, and thermal isolation, while re-applying design and technology solutions that have been demonstrated for X-IFU. Figure 8 shows the current mechanical implementation of the focal plane.

Background-limited KID detectors optimized to meet PRIMAger's performance requirements are sensitive to power loading from both out-of-band radiation and straylight incident on the detectors and modulation of the detector response by quasi-static magnetic fields in the detectors' environment. This is why the design include optical bandpass filtering and straylight control within the focal planes. In addition, a "clean" magnetic environment that is needed to operate the KID detectors with their required performance is realized by a two-layer magnetic shield, with a cryogenic mu-metal (CMM) shield attenuating the (static) field that is present in the FPMA environment at cooldown by a factor of ~50 and a superconducting (Nb) shield combining with the CMM shield to provide a shielding factor of > $2x10^3$ relative to time-varying fields in the FPMA environment during operation.

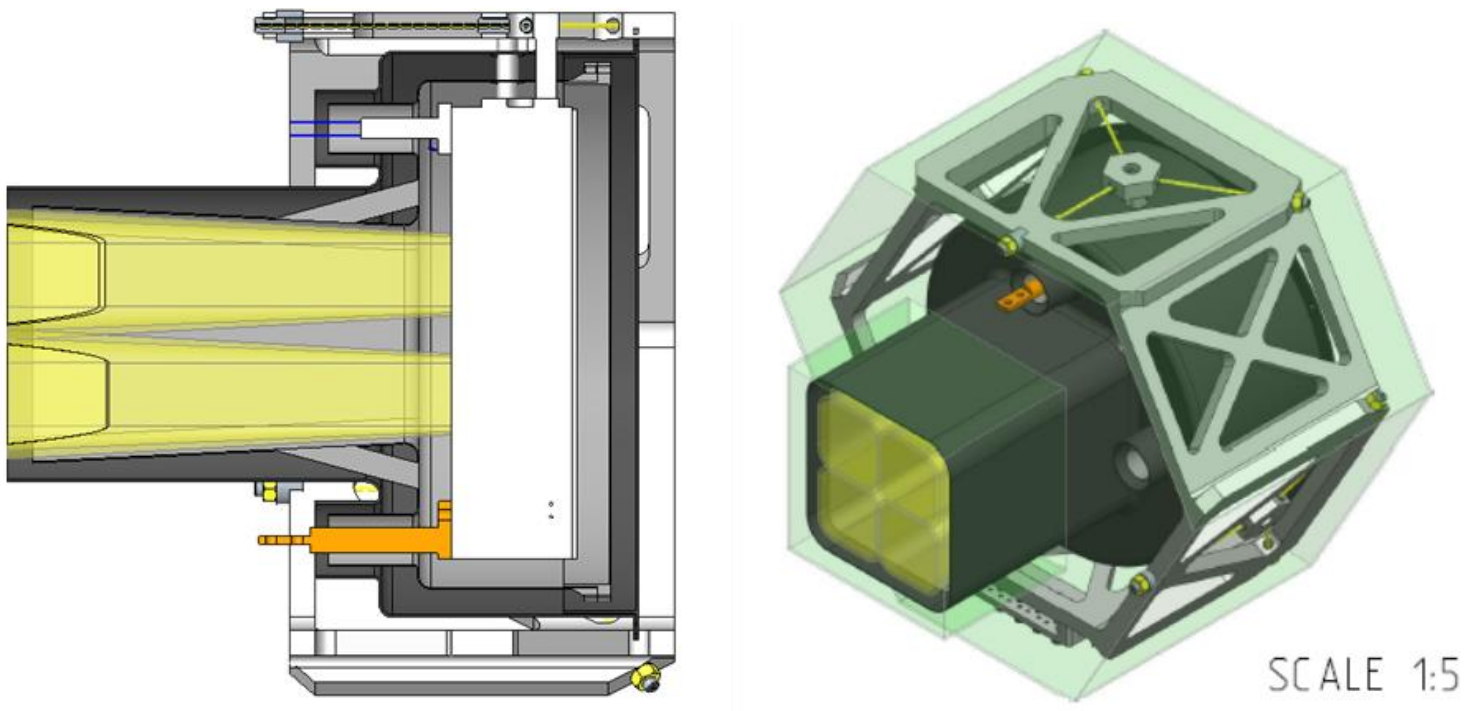

*Figure 8. Preliminary PPI FPMA design concept (left). A cut-away view with the T0 stage (white) suspended inside a 2-layer magnetic shield (Nb shield in light grey and CMM shield in dark grey) and an "open" T1 structure. The yellow "cones" propagating from the T0 stage are illustrative of the optical beams incident on the 4 PPI detectors. (right) External view.*

The mapping of the detector readout channels to the PHI and PPI detector arrays is driven by a combination of:

- the allocation of 4 detector readout channels to each of the PPI and PHI FPMAs;
- the maximum multiplexing factor of 500 pixels per readout channel;
- the requirement that any one readout channel not read-out more than ½ of any one detector array; and
- the PHI and PPI detector array sizes.

In particular, with ~50% of the 1949 PPI pixels found in PPI1 and ~75% of the 1940 PHI pixels found in PHI1, this set of requirements is met by the channel-to-array mapping illustrated in Fig. 9.

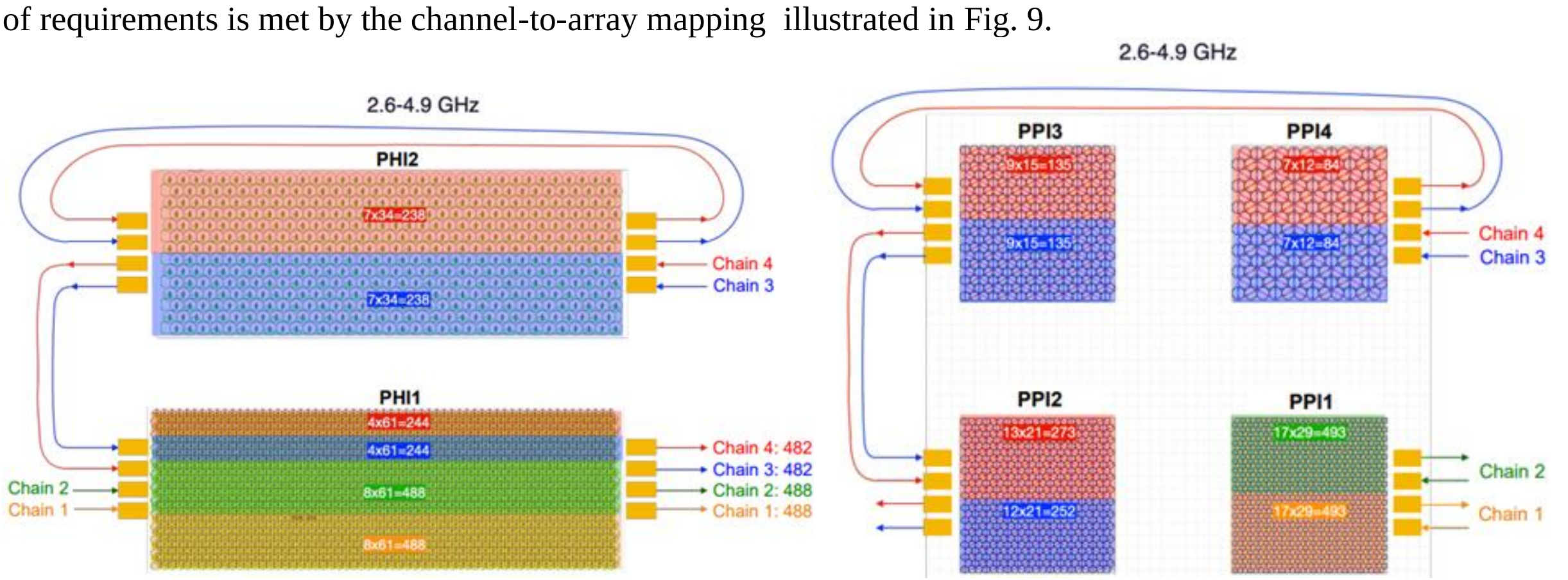

*Figure 9. Schematic representation of the PRIMAger detector to readout mapping for the PHI (left) and PPI (right) focal planes*

## 3.3 TLA description

The Low Temperature Systems Department of the Alternative Energies and Atomic Energy Commission Commissariat (CEA-DSBT) is responsible for the design and fabrication of the Thermal Link Assembly. The TLA provides a single 0.12 K interface for the two PRIMAger FPMAs. It damps vibrations and thermally isolates the 0.12 K thermal strap from its 1 K mount with a Kevlar cord system designed for the GEVS 100 g static load. The high-conductivity copper bus from the TLA to the FPMs provides a thermal conductance >5 µW/mK at 0.12 K while minimizing mechanical coupling. This link, combined with the two FPMA heat capacities, provides a short thermal response time below the 2.5 s requirement, enabling precise detector temperature control by the CADR. The 1 K thermal link from the CADR connects directly to the MOTA and provides a cooling power of 114 µW. Thermal modeling shows a low thermal gradient <0.02 K in the MOTA all-aluminum structure.

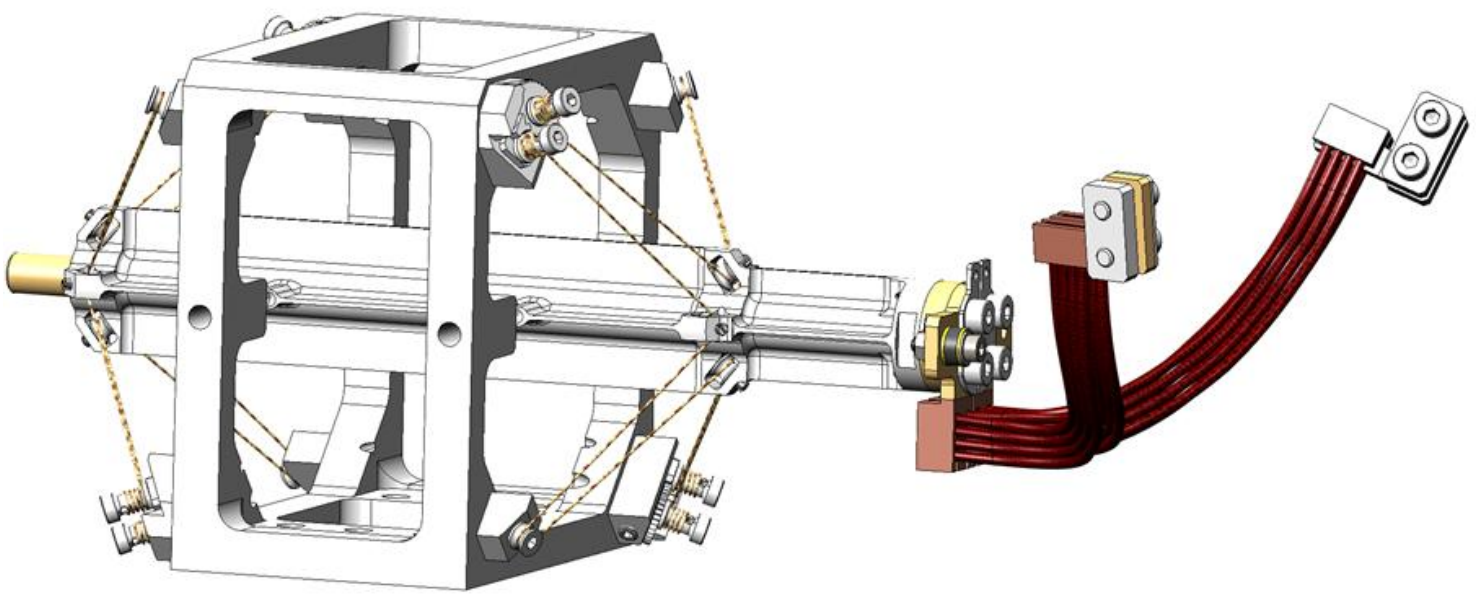

*Figure 10. TLA CAD*

## 3.4 Filter assembly

CTT/Cardiff University is providing the quasi-optical filters for PRIMAger except the LVF assembly that is provided by SRON. CTT/Cardiff Univ. is responsible for the design, manufacturing, test and delivery of the filters for MOTA and FPMA. A preliminary design is proposed by CTT for the filters, that are based on a precisely spaced multiple copper meshes embedded in transparent polypropylene. The current proposed design is to have 3 filters within the MOTA optical path, which are a series of low-pass filters, and several filters within FPMA that will provide in addition band pass capacity for PHI and PPI. The rejection factors required by PRIMAger are provided in the following table. It is expected that the transmission of the full assembly of quasioptical filter will be better than 0.53 for PPI and 0.62 for PHI, without the LVF.

*Tableau 1. Transmission on the different band. For PHI1 channel, λ1 = 24µm and for PHI2 λ2 =45µm. Each PPI channel shall be considered for the band for PPI.*

| Band (µm) | PPI | PHI |
|---|---|---|
| 0.1 - 3 | < 5e-13 | < 5e-13 |
| 3 - 24 | < 2e-12 | < 2e-12 |
| 24 – Lambda 1 | < 3e-8 | < 1e-5 |
| Lambda 1 – Lambda 2 | 0.53 | 0.62<br>0.6 for LVF assembly |
| Lambda2 - 235 | < 3e-2 | < 1e-3 |
| 235 - 700 | < 3e-3 | < 1e-3 |
| 700 - 3300 | < 1e-5 | < 1e-5 |

# 4. PRIMAGER INSTRUMENT DEVELOPMENT PLAN

## 4.1 PRIMAger development approach

The verification process and development for PRIMAger will be implemented at different verification stages to minimize project risk. We propose the following model approach, described hereafter in more details:

- Instrument models to analyse design performance based on CAD model, ZEMAX model and a thermal model (Nodal ESATAN ThermXL and then FEM as thermal desktop)
- Prototypes development at early stage of the project to demonstrate design concept and feasibility. Goal is to demonstrate maturity level of TRL6 at least. The prototypes developed during phase A are:
  - A TLA prototype, by CEA DBST
  - A monopod prototype for MOTA and a MOTA bipod prototype with representative mass of sub-systems, by LAM
  - A MOTA mirror prototype and a mounted mirror, by LAM
  - A PHI detector TRL6
- A FPMA Engineering Model (EM) for PPI detector, by SRON and a TLA Engineering Model (EM), by CEA DBST. They will be tested at qualification level to demonstrate the qualification and compliance of the design and the performance, with the the specified requirements at specified levels.

- PRIMAger Cold Unit Proto Flight Model (PFM) including MOTA, FPMA, TLA, ELA and FA PFMs, tested at acceptance level, to ensure that the product is in agreement with the qualified design.

It is foreseen a partially parallel development of the detector EM and the detectors FM to optimise the schedule. Several spare parts will also be manufactured. Due to the tight schedule, it has been decided at the early stage of the project to not develop a Structural and Thermal Model (STM) of the instrument nor an Engineering and Qualification Model (EQM). The demonstration of the structural and thermal performance is foreseen through the prototypes development first and then on the PFM sub-systems.

The PRIMAger integration and test flow will then be:

- Integration and test of the sub-systems at institutes premises (NOTA: integration of the FPMA filters and JPL thermal sensors in FPMA are done during FPMA AIT)
- Integration of the MOTA filter in MOTA
- Integration of TLA on MOTA
- Integration of FPMA on MOTA
- Integration of ELA on MOTA
- Test of PRIMAger Cold unit with EGSE provided by NASA in the TVOGSE

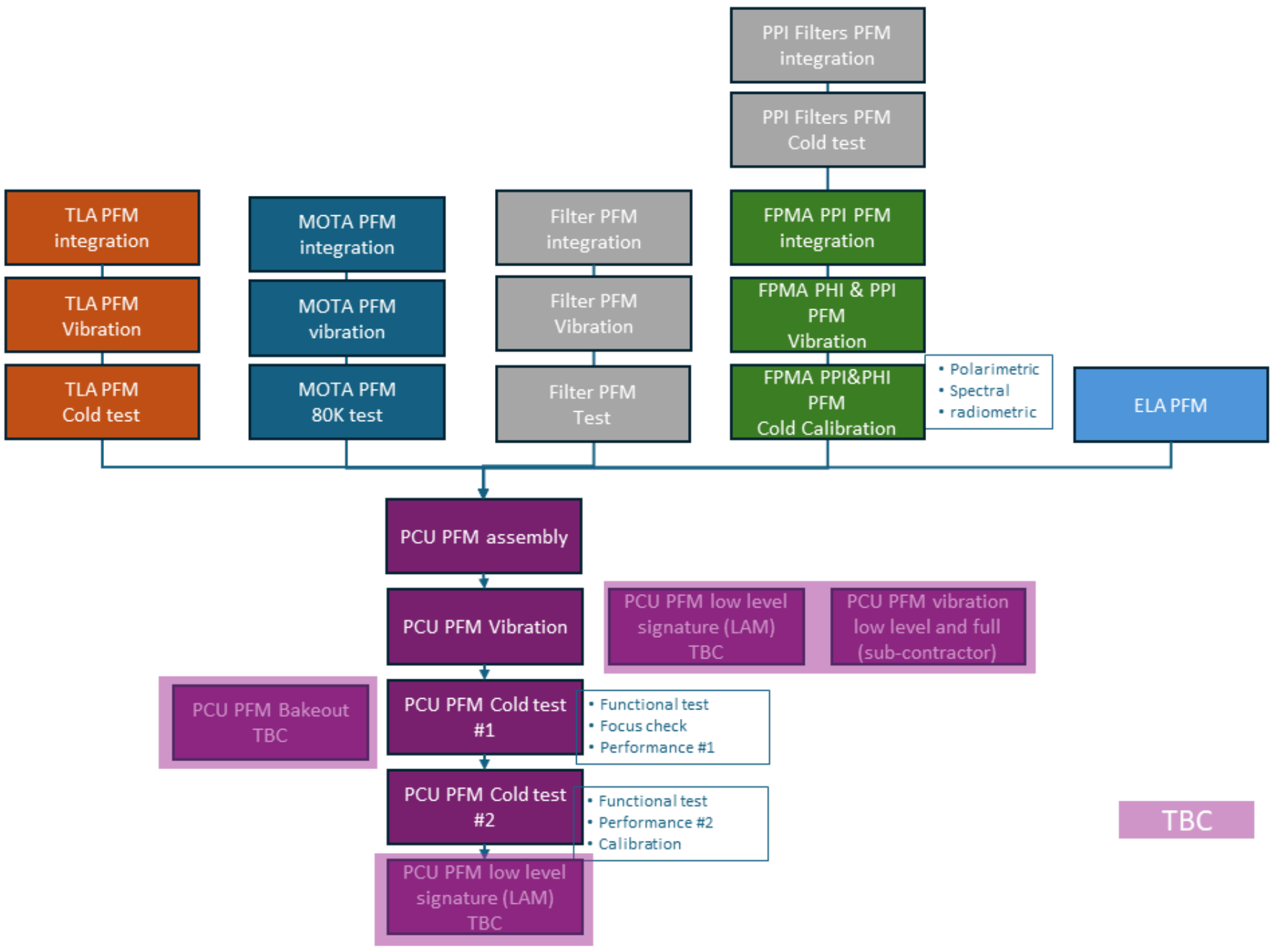

*Figure 11. Integration and test flow for PRIMAger.*

## 4.2 Verification approach

Figure 12 shows the verification strategy in more details that we propose for PCU at the different levels and development steps. The main philosophy is to test as much as possible the performance of the detectors at detectors (FPMA) level. During the TBTV test campaign, we will recheck some of the parameters of the detectors to make sure that the integration on the instrument has no impact. The main test at PCU level will be a sensitivity measurement.

We will then have a TBTV test campaign with different goals:

- Optical performance verification, in particular co-focus between PHI and PPI and PSF shape, FoV measurements
- Measurement of PHI and PPI sensitivity in different background and flux conditions
- Validation of the thermal performance of the instrument design. We will verify the thermal budget and all associated thermal performance requirements
- Calibration of the instrument and in particular its polarimetric signature. The polarisation calibration on ground will be done both at component level, to build an accurate model of the instrument, and at PCU level to compare the measurements on ground with the expected values from the model. This will allow to validate the model we will use during flight and operation of PRIMA.

The verification plan for PCU can be summarized as:

- A sequence of performance test, mostly linked to optical performance and to spectral and polarisation calibration.
- A sequence of thermal test
- A sequence of mechanical performance test, such as vibration

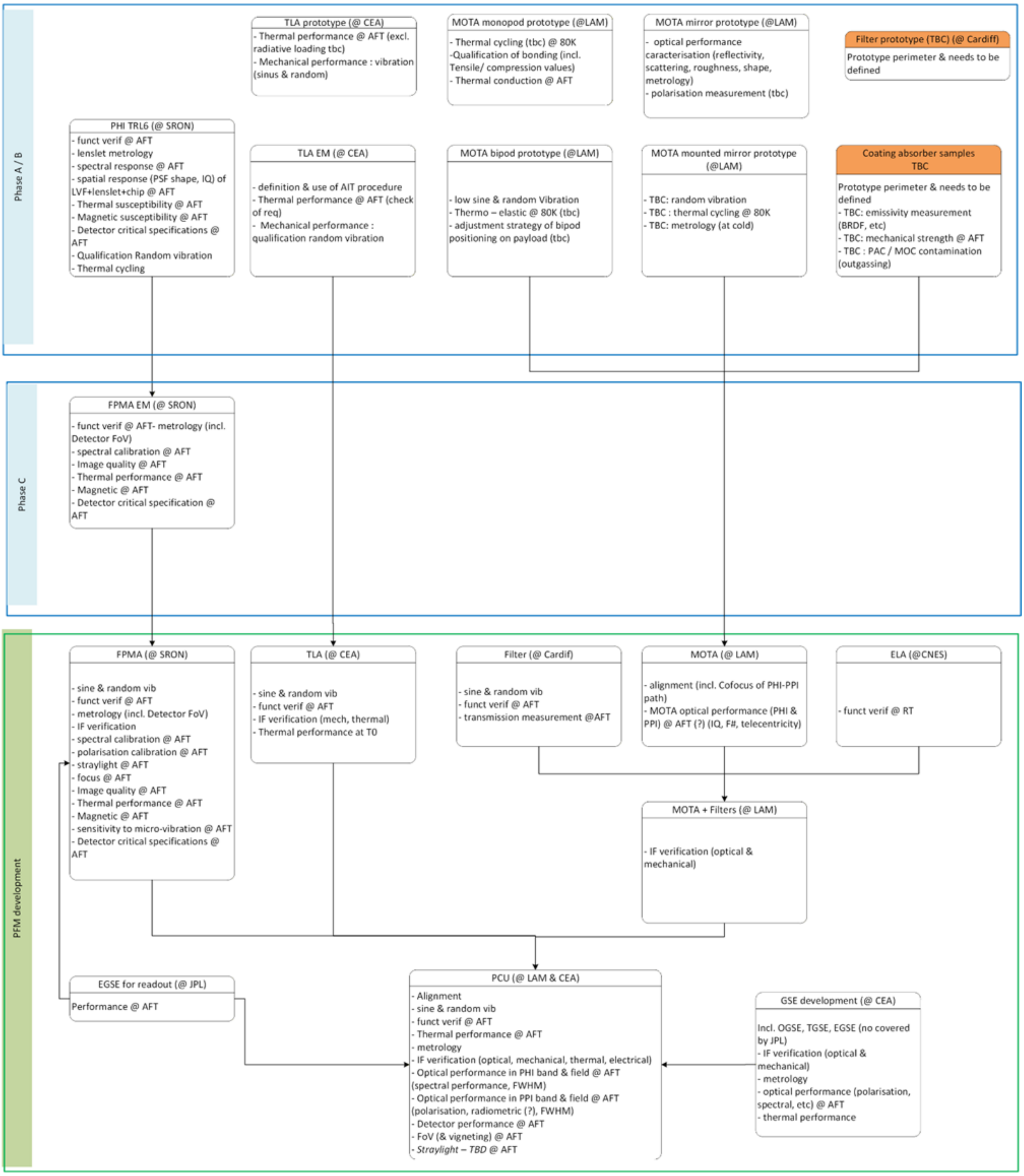

*Figure 12. Verification strategy for PRIMAger.*

## 5. CONCLUSION

PRIMAger will be a key instrument on board of the PRIMA mission developed by NASA. It will provide both an hyperspectral imager and a polarimetric imager to address fundamental astrophysical questions and expand observational capability in the far IF domain. The design of the instrument has been developed during the selection phase of NASA. The main architecture and design trade-off have been made to ensure a design within the expected performance. Design is progressing in order to be ready for the start of the phase B foreseen by end of 2026, while the verification strategy of the instrument is being developed.